\documentclass[11pt,a4paper]{article}

\usepackage[utf8]{inputenc} 
\usepackage[T1]{fontenc}    %
\usepackage[english]{babel} 
\usepackage[final]{microtype} 

\usepackage{amssymb,amsmath,mathtools} 
\usepackage{graphicx} 
\usepackage{bbm} 
\usepackage{tikz} 

\usepackage[hmargin=0.12\paperwidth,vmargin=0.2\paperwidth,bindingoffset=0.5cm]{geometry}

\newcommand{\qi}{\mathbbm i} 
\newcommand{\qj}{\mathbbm j}
\newcommand{\qk}{\mathbbm k}

\newcommand{\p}{\partial} 

\newcommand{\ph}{\hphantom} 

\newcommand{\nn}{\nonumber} 

\newcommand{\R}{\mathbb{R}} 
\newcommand{\C}{\mathbb{C}}

\newcommand{\cO}{\mathcal O}
\newcommand{\cC}{\mathcal C}

\newcommand{\cP}{\mathcal P}

\let\olddet\det
\renewcommand{\det}{\olddet\nolimits} 

\renewcommand{\phi}{\varphi} 
\renewcommand{\epsilon}{\varepsilon} 

\DeclareMathOperator{\tr}{Tr} 
\DeclareMathOperator*{\pf}{Pf} 
\DeclareMathOperator*{\Qdet}{Qdet} 
\DeclareMathOperator*{\dof}{d.o.f.} 
\DeclareMathOperator{\erfc}{erfc} 

\DeclarePairedDelimiter{\abs}{\lvert}{\rvert} 
\DeclarePairedDelimiter{\average}{\langle}{\rangle} 
\DeclarePairedDelimiter{\inner}{\langle}{\rangle} 

\newcommand{\jpdf}{\text{jpdf}}
\newcommand{\macro}{\text{macro}}
\newcommand{\bulk}{\text{bulk}}
\newcommand{\edge}{\text{edge}}
\newcommand{\origin}{\text{origin}}

\newcommand{\MeijerG}[8][\bigg]{G^{{ #2 },\,{ #3 }}_{{ #4 },\,{ #5 }} #1( \begin{matrix} #6 \\ #7 \end{matrix}\, #1\vert\, #8 #1)}
\newcommand{\hypergeometric}[6][\bigg]{\,{}_{#2} F_{#3} #1( \begin{matrix} #4 \\ #5 \end{matrix}\, #1\vert\, #6 #1)}

\usepackage{hyperref}   
\hypersetup{
pdfborder={0 0 0},      
pdfauthor={J. R. Ipsen} 
}

\title{Products of Independent Quaternion Ginibre Matrices\\ and their Correlation Functions}
\author{J.~R.~Ipsen\\
\small Department of Physics, Bielefeld University,\\
\small Postfach 100131, D-33501 Bielefeld, Germany}
\date{\today}

\begin{document}

\maketitle

\begin{abstract}
\noindent
We discuss the product of independent induced quaternion ($\beta=4$) Ginibre matrices, and the eigenvalue correlations of this product matrix. The joint probability density function for the eigenvalues of the product matrix is shown to be identical to that of a single Ginibre matrix, but with a more complicated weight function. We find the skew-orthogonal polynomials corresponding to the weight function of the product matrix, and use the method of skew-orthogonal polynomials to compute the eigenvalue correlation functions for product matrices of finite size. The radial behavior of the density of states is studied in the limit of large matrices, and the macroscopic density is discussed. The microscopic limit at the origin, at the edge(s) and in the bulk is also discussed for the radial behavior of the density of states.
\end{abstract}

\newpage

\section{Introduction}
\label{sec:intro}

Random matrix theory has found a vast number of applications in physics and mathematics as well as many other sciences; we refer to~\cite{ABF:2011} for a recent compilation. But in certain applications it is insufficient to consider a single random matrix, and one must consider products of random matrices instead. Two limits are particular interesting: Products of a large number of matrices, and products of large matrices. As applications, which belong to the first category, we may mention: Lyapunov exponents for disordered and chaotic systems~\cite{CPV:1993}, multiplicative matrix-valued non-commutative diffusion processes~\cite{JLJN:2002}, stability of ecological systems~\cite{Caswell:2001}, and scattering of electromagnetic waves on random obstacles in telecommunication~\cite{TV:2004}. These have all been studied using products of random matrices. The limit of large matrices has been used to study: Transfer matrices in mesoscopic wires~\cite{Beenakker:1997}, large $N_c$ Wilson loops in Yang--Mills 
theory~\cite{BLNT:2005,NN:2007,JW:2004,BN:2008}, and quantum chromodynamics (QCD) at finite chemical potential~\cite{Akemann:2007}.

Recently, several papers have appeared which discuss the product of large non-Hermitian matrices. The macroscopic, or mean, density of states for complex non-Hermitian matrices has been discussed using diagrammatic methods in~\cite{GJJN:2003,BJW:2010,BJLNS:2010}. Proofs for the macroscopic behavior for products of complex non-Hermitian matrices can be found in~\cite{GT:2010,RS:2011}. The method of orthogonal polynomials was used in~\cite{AB:2012} to obtain explicit expressions for the eigenvalue correlation functions for a product of independent complex ($\beta=2$) Ginibre matrices of finite size, and it was shown that the macroscopic limit agreed with previous results. A discussion of the microscopic behavior at the origin and at the edge as well as in the bulk of the spectrum was also presented in~\cite{AB:2012}, while hole probability and overcrowding at the origin was discussed in~\cite{AS:2012}.

The product of two independent non-Hermitian matrices has been discussed for all three Ginibre ensembles: The real ($\beta=1$) Ginibre ensemble was discussed in~\cite{APS:2009,APS:2010,KWY:2010,AKPW:2011}, the complex ($\beta=2$) Ginibre ensemble in~\cite{Osborn:2004,KS:2010}, and the quaternion ($\beta=4$) Ginibre ensemble in~\cite{Akemann:2005}. But to our knowledge, the product of three or more Ginibre matrices has only been discussed in the complex case. It is desirable task to extent this discussion to the other ensembles: real ($\beta=1$) and quaternion ($\beta=4$) matrices.

In this paper we discuss the statistical properties of the eigenvalues of a product of quaternion ($\beta=4$) Ginibre matrices, which is a natural extension of the discussions of single quaternion Ginibre matrix~\cite{Ginibre:1965,Kanzieper:2002} as well as the discussion of the product of complex ($\beta=2$) Ginibre matrices~\cite{AB:2012}. As mentioned above, the product of random matrices have a broad range of applications. However, quaternion matrices are less studied and fewer applications are known. As an explicit example of an application for the product of quaternion Ginibre matrices, we can mention that the product of two quaternion Ginibre matrices is directly related to QCD with a finite chemical potential and fermions in the adjoint representation, or two-color QCD with staggered lattice fermions in the fundamental representation~\cite{Akemann:2005}. Here, we will extent the discussion to the product of any finite number of matrices, which at present is mostly of mathematical interest.

We will use the method of skew-orthogonal polynomials to obtain explicit expressions for the eigenvalue correlation functions for matrices of finite size. We will also discuss the radial behavior for the density of states in the limit of large matrices. This allows us to find the macroscopic density as well as microscopic limits at the origin, at the edge and in the bulk.

We consider the product matrix
\begin{equation}
P_n^m=X_1X_2\cdots X_n,
\label{intro:product}
\end{equation}
where $X_i$ are independent matrices drawn from the induced quaternion ($\beta=4$) Ginibre ensemble, specified by the joint probability density function
\begin{equation}
p_\jpdf^m(X)=C_N^m \det (X)^m\exp[-\tr X^\dagger X],
\label{intro:pdf}
\end{equation}
where $m$ is a non-negative number and $C_N^m$ is a normalization constant. Note that we look at the induced Ginibre ensemble (if $m\neq 0$); a similar generalization could be introduced in the complex case~\cite{AB:2012} without much difficulty. Looking at the induced Ginibre ensemble allow us to introduce an inner edge for the spectrum, such that the eigenvalues are located within an annulus rather than a disk, as we will discuss more thoroughly in sections~\ref{sec:radial} and~\ref{sec:micro}. The inner edge introduced by the parameter $m$ was studied for a real ($\beta=1$) and a complex ($\beta=2$) induced Ginibre matrix in~\cite{FBKSZ:2012}. That this situation is particular interesting is illustrated by the single ring theorem~\cite{FZ:1997,GKZ:2011}. It should mentioned that the single ring theorem is directly related to the Haagerup--Larsen theorem~\cite{HL:2000}.

Here and in the following $X=[q_{ij}]$ denotes a non-Hermitian $N\times N$ matrix with independent quaternion entries
\begin{equation}
q=q^{(0)}+\qi q^{(1)}+\qj q^{(2)}+\qk q^{(3)},\quad q^{(k)}\in\R.
\end{equation}
Equivalently, we may represent the quaternion units as $2\times 2$ matrices
\begin{equation}
\qi=\begin{pmatrix} i & 0 \\ 0  & -i \end{pmatrix},\quad
\qj=\begin{pmatrix} 0 & 1 \\ -1 & 0  \end{pmatrix}\quad\text{and}\quad
\qk=\begin{pmatrix} 0 & i \\ i  & 0  \end{pmatrix},
\end{equation}
where $i$ is the imaginary unit. In this representation $X=[q_{ij}]$ is a $2N\times 2N$ matrix with a block structure given by
\begin{equation}
q=\begin{pmatrix} u & -v^\ast \\ v  & u^\ast \end{pmatrix},\quad u,v\in\C.
\label{intro:q}
\end{equation}
Note that the determinant and the trace in the probability density function~\eqref{intro:pdf} should be understood as applied to this $2N\times 2N$ matrix. Moreover, it should be noted that the eigenvalues come in complex conjugate pairs such that the determinant in~\eqref{intro:pdf}, and therefore the entire density function, is real and positive. For a more thorough discussion of quaternion matrices see e.g.~\cite{Mehta:2004}. 

The partition function corresponding to the product matrix~\eqref{intro:product} is given by
\begin{equation}
Z^{n,m}_N=\int \abs{DX}\prod_{i=1}^n p_\jpdf^m(X_i)=\int \abs{DX}\prod_{i=1}^nC_N^m\det (X_i)^m \exp[-\tr X_i^\dagger X_i].
\label{intro:Z}
\end{equation}
 In our notation $DX$ denotes the Euclidean volume element (i.e. the exterior product of all independent one-forms $DX=\bigwedge_\alpha dX_\alpha$) and $\abs{DX}$ is the corresponding unoriented volume element. This means that
\begin{equation}
DX=\bigwedge_{i=1}^n DX_i=\bigwedge_{i=1}^n \bigwedge_{a,b} (dX_i)_{ab}=\bigwedge_{i=1}^n \bigwedge_{a,b} \bigwedge_{q=0}^3 (dX_i^{(q)})_{ab},
\end{equation}
where $i$ is the index for the $n$ independent matrices, $ab$ are the indices for the $N^2$ independent matrix entries, and $q$ is the quaternion index. The corresponding unoriented volume element is
\begin{equation}
\abs{DX}=\prod_{i=1}^n \prod_{a,b} \prod_{q=0}^3 (dX_i^{(q)})_{ab}.
\end{equation}
The main interest of this paper is twofold: First, we want to rewrite the partition function~\eqref{intro:Z} in terms of the eigenvalues of product matrix $P_n^m$, i.e. replace the joint probability density functions, $p_\jpdf^m$ for the individual matrices, $X_i$, with a joint probability density function, $\cP_\jpdf^{n,m}$, for the eigenvalues of the product matrix, $P_n^m$. Second, we want to use this insight to study the eigenvalue correlations of the product matrix.

The paper is organized as follows: In section~\ref{sec:jpdf} we discuss how to obtain the joint probability density function, $\cP_\jpdf^{n,m}$, for the eigenvalues of the product matrix, $P_n^m$. Section~\ref{sec:sop} explains how to find the eigenvalue correlations of the product matrix, $P_n^m$, using the method of skew-orthogonal polynomials. The large-$N$ behavior of the radial behavior for the density of states of the product matrix is discussed in sections~\ref{sec:radial} and~\ref{sec:micro}. Section~\ref{sec:radial} focuses on the macroscopic behavior, while section~\ref{sec:micro} discusses the microscopic limits. Some technical details are given in the appendices. 

\section{Joint Probability Density Function}
\label{sec:jpdf}

In this section we outline the derivation of the joint probability density function, $\cP_\jpdf^{n,m}$, for the eigenvalues of the product matrix $P_n^m$. We triangularize the independent Ginibre matrices, $X_i$, and use this parameterization to obtain the joint probability density function for the eigenvalues of the product matrix. Some technical details are given in appendix~\ref{sec:jacob}.

We parameterize the quaternion matrices $X_i$ (and therefore also product matrix $P_n^m$) using a generalized Schur decomposition
\begin{equation}
X_i=U_i(\Lambda_i+T_i)U_{i+1}^{-1}.
\end{equation}
Here $\Lambda_i$ are diagonal matrices, $T_i$ are strictly upper-triangular matrices and $U_i$ are symplectic matrices with $U_{n+1}=U_1$. This parameterization was used to study the $n=2$ case in~\cite{Akemann:2005}, and a similar parameterization has been used to discuss the product of $n$ independent complex ($\beta=2$) Ginibre matrices~\cite{AB:2012}. The product of two complex Ginibre matrices was first discussed in~\cite{Osborn:2004,KS:2010}. Also the product of two independent real ($\beta=1$) Ginibre matrices have been discussed in the literature~\cite{APS:2009,APS:2010,AKPW:2011}.

In the quaternion case, each $\Lambda_i$ is a diagonal (block) matrix, where the diagonal entries are given by
\begin{equation}
(\Lambda_i)_{aa}=\lambda_{ia}^{(0)}+\qi\lambda_{ia}^{(1)}=\begin{pmatrix} x_{ia} & 0 \\ 0 & x_{ia}^\ast \end{pmatrix}, 
\quad\text{where}\quad
x_{ia}=\lambda_{ia}^{(0)}+i\lambda_{ia}^{(1)}\in\C;
\end{equation}
each $T_i$ is a strictly upper-triangular (block) matrix, e.g. $(T_i)_{ab}=0$ for $a\geq b$ and $(T_i)_{ab}$ for $a<b$ are given by the quaternion structure~\eqref{intro:q}; and each $U_i$ is a symplectic matrix. In order to ensure that the parameterization is unique it is necessary to introduce further constraints on the symplectic matrices, $U_i$. We see that the structure of each triangular matrix, $\Lambda_i+T_i$, is invariant under transformations
\begin{equation}
U_i\mapsto U_iV_i
\end{equation}
where $V_i$ is a diagonal matrix, where the diagonal elements are $2\times 2$ diagonal unitary matrices. We choose $U_i\in Sp(N)/U(1)^N$ such that the parameterization is unique. The degrees of freedom ($\dof$) are (obviously) conserved by the parameterization,
\begin{equation}
\dof_{i=1,\ldots,n} (X_i)=\dof_{i=1,\ldots,n} (\Lambda_i,T_i,U_i).
\end{equation}
It is well-known that for $n=1$ the Jacobian corresponding to this change of variables is given by~\cite{Ginibre:1965}
\begin{equation}
\abs{J(z)}=\prod_{a>b}^N\, \abs{z_b-z_a}^2 \abs{z_b-z_a^\ast}^2 \prod_{c=1}^N\, \abs{z_c-z_c^\ast}^2,
\end{equation}
where $z_k$ are the eigenvalues of $X_1=P_{n=1}^m$. The derivation of the Jacobian for arbitrary $n$ is given in appendix~\ref{sec:jacob}. The generalized Jacobian has a structure similar to the $n=1$ case, except that the Jacobian now depends on the product of eigenvalues of the individual matrices $X_i$, i.e. $J(\prod_{i=1}^n x_{ia})$. Moreover, the joint probability density functions~\eqref{intro:pdf} for the individual matrices $X_i$ are independent of the symplectic matrices, $U_i$, such that the integration over these degrees of freedom only contribute to the normalization constant. Also the integration over the triangular matrices, $T_i$, contribute only to the normalization; here the exponential term, $\exp[-\tr T_i^\dagger T_i]$, ensures that the integral is finite. For this reason we may write the partition function~\eqref{intro:Z} in terms of the eigenvalues, $z_k$, of the product matrix~\eqref{intro:product},
\begin{equation}
Z_N^{n,m}=\prod_{k=1}^N\int_\C d^2z_k\,\cP_\jpdf^{n,m}(z_1,\ldots,z_N),
\label{jpdf:Z}
\end{equation}
and the joint probability density function for the product matrix is given by
\begin{equation}
\cP_\jpdf^{n,m}(z_1,\ldots,z_N)
=\cC_N^{n,m} \prod_{c=1}^N w_n^m(z_c)\, \abs{z_c-z_c^\ast}^2 \prod_{a>b}^N\, \abs{z_b-z_a}^2 \abs{z_b-z_a^\ast}^2 ,
\label{jpdf:jpdf}
\end{equation}
where $\cC_N^{n,m}$ is a normalization constant. We take the normalization to be $\cC_N^{n,m}=1/4$, which ensures that the density of states is normalized to the number of eigenvalues as we will see below. The weight function, $w_n^m(z)$, is defined using a Dirac delta function in order to write the joint probability function~\eqref{jpdf:jpdf} as a function of the eigenvalues of the product matrix rather than a function of the eigenvalues of the individual matrices, $X_i$, 
\begin{equation}
w_n^m(z)\equiv\prod_{i=1}^n\int_\C d^2x_i\abs{x_i}^{2m}e^{-\abs{x_i}^2}\delta^2(z-x_1x_2\cdots x_n).
\label{jpdf:weight1}
\end{equation}
For $m=0$ this weight function is identical to the weight discussed in~\cite{AB:2012}, and here it was shown that the weight function could be expressed as a Meijer $G$-function. It turns out that the induced weight function~\eqref{jpdf:weight1} also can be expressed as a Meijer $G$-function,
\begin{equation}
w_n^m(z)=\pi^{n-1}\MeijerG{n}{0}{0}{n}{-}{m,\ldots,m}{\abs z^2}.
\label{jpdf:weight2}
\end{equation}
The weight~\eqref{jpdf:weight2} can be found following the same idea as in~\cite{AB:2012}. The main idea is to find a recursion relation for the Mellin transform of the family of weights $\{w_n^m\}_{n\geq 1}$ given by~\eqref{jpdf:weight1}. Solving the recursion relation and taking the inverse Mellin transform yields the weight function given in equation~\eqref{jpdf:weight2}. But since we already know the weight function for $m=0$, we may take a shortcut. We use the fact that
\begin{equation}
\prod_{i=1}^n\det(X_i)^m=\det (P_n^m)=\prod_{k=1}^N\abs{z_k}^2,
\end{equation}
where $z_k$ (and $z_k^\ast$) are the eigenvalues of the product matrix $P_n^m$ defined in~\eqref{intro:product}. It follows that the weight with $m\neq 0$ is related to weight with $m=0$ by
\begin{equation}
w_n^m(z)=\abs{z}^{2m}w_n^{m=0}(z).
\end{equation}
It follows directly from the definition of the Meijer $G$-function~\cite{GR:2000} that the induced weight may be written as~\eqref{jpdf:weight2}.

\section{Correlations and Skew-Orthogonal Polynomials}
\label{sec:sop}

One is often interested in the $k$-point correlation functions for the eigenvalues,
\begin{equation}
R_k^{n,m}(z_1,\ldots,z_k)=\frac{N!}{(N-k)!}\prod_{h=k+1}^N\int_\C d^2z_h\,\cP_\jpdf^{n,m}(z_1,\ldots,z_N),
\label{sop:R}
\end{equation}
which loosely speaking give the probability of finding $k$ eigenvalues at the positions $z_1$,\,\ldots,$z_k$ in the complex plane. In this section we will find an explicit expression for the eigenvalue correlations~\eqref{sop:R} using the method of skew-orthogonal polynomials.

Following~\cite{Kanzieper:2002}, we use a standard identity for determinants to write the joint probability density function~\eqref{jpdf:jpdf} as
\begin{equation}
\cP_\jpdf^{n,m}(z_1,\ldots,z_N) =\frac{1}{4} \prod_{h=1}^N w_n^m(z_h)\,(z_h^\ast-z_h)
\olddet_{\substack{\ell=1,\ldots,N \\ k=1,\ldots,2N}} \begin{bmatrix} z_\ell^{k-1} \\ z_\ell^{\ast k-1}\end{bmatrix}.
\end{equation}
As usual we rewrite the determinants in terms monic polynomials, and it follows directly from de~Bruijn's integration theorem~\cite{deBruijn:1955} that the partition function~\eqref{jpdf:Z} may be written as a Pfaffian,
\begin{equation}
Z_N^{n,m}=(2N)!\pf_{1\leq k,\ell\leq 2N}\Big[ \frac{1}{4} \int_\C d^2z\, w_n^m(z)(z^\ast-z) 
\big[p_{k-1}^{n,m}(z)p_{\ell-1}^{n,m}(z^\ast)-p_{k-1}^{n,m}(z^\ast)p_{\ell-1}^{n,m}(z)\big] \Big],
\label{sop:Z}
\end{equation}
where $p_k^{n,m}(z)$ are monic polynomials of order $k$. It is clear that the expression~\eqref{sop:Z} for the partition function simplifies considerably if we choose the polynomials, $p_k^{n,m}(z)$, to be skew-orthogonal with respect to the skew-product
\begin{equation}
\inner{f\vert g}_S=\frac{1}{4}\int_\C d^2z\, w_n^m(z)(z^\ast-z) \big[f(z)g(z^\ast)-f(z^\ast)g(z)\big].
\label{sop:inner}
\end{equation}
Explicitly, the skew-orthogonality means that
\begin{subequations}
\label{sop:rel}
\begin{align}
\inner{p_{2k+1}^{n,m}\vert p_{2\ell+1}^{n,m}}_S &= \inner{p_{2k}^{n,m}\vert p_{2\ell}^{n,m}}_S=0, \label{sop:rel1}\\
\inner{p_{2k+1}^{n,m}\vert p_{2\ell}^{n,m}}_S   &= h_k^{n,m}\delta_{k\ell}, \label{sop:rel2}
\end{align}
\end{subequations}
where $h_k^{n,m}$ are constants. Using the skew-orthogonality of the polynomials we may write the partition function~\eqref{sop:Z} as~\cite{Mehta:2004}
\begin{equation}
Z_N^{n,m}=(2N)!\, \prod_{k=0}^{N-1} h_k^{n,m}.
\end{equation}
In the following we will always assume that the polynomials, $p_k^{n,m}(z)$, are skew-orthogonal, i.e. that the polynomials satisfy the skew-orthogonality conditions~\eqref{sop:rel}. We will find the explicit realization of the polynomials, which are skew-orthogonal with respect to~\eqref{sop:inner}, at the end of this section.

The skew-orthogonal polynomials are extremely useful, and not only the partition function~\eqref{sop:Z} but also the eigenvalue correlations~\eqref{sop:R} can be expressed in terms of the skew-orthogonal polynomials. One derivation of the eigenvalue correlations is given in~\cite{Kanzieper:2002}, which uses an idea presented in~\cite{TW:1998}. The eigenvalue correlations may be written as
\begin{equation}
R_k^{n,m}(z_1,\ldots,z_k)=\prod_{h=1}^k w_n^m(z_h)(z_h^\ast-z_h)\pf_{1\leq i,j\leq k}
\label{sop:R1}
\begin{bmatrix}
\kappa_N^{n,m}(z_i,z_j^\ast) & \kappa_N^{n,m}(z_j^\ast,z_i^\ast) \\
\kappa_N^{n,m}(z_i,z_j)      & \kappa_N^{n,m}(z_j^\ast,z_i)
\end{bmatrix},
\end{equation}
where the so-called prekernel $\kappa_N^{n,m}(u,v)$ may be expressed in terms of the skew-orthogonal polynomials,
\begin{equation}
\kappa_N^{n,m}(u,v)=\sum_{k=0}^{N-1}\frac{p_{2k+1}^{n,m}(u)p_{2k}^{n,m}(v)-p_{2k+1}^{n,m}(v)p_{2k}^{n,m}(u)}{h_k^{n,m}}.
\label{sop:ker1}
\end{equation}
In the literature one sometimes uses the notation of quaternion determinants, see e.g.~\cite{Kanzieper:2002,Akemann:2005}. In this notation the eigenvalue correlations are written as
\begin{equation}
R_k^{n,m}(z_1,\ldots,z_k)=\prod_{h=1}^k w_n^m(z_h)(z_h^\ast-z_h) \Qdet_{1\leq i,j\leq k} K_N^{n,m}(z_i,z_j),
\label{sop:R2}
\end{equation}
where $\Qdet$ is the quaternion determinant and $K_N^{n,m}(u,v)$ is a $2\times 2$ matrix kernel. The relation between~\eqref{sop:R1} and~\eqref{sop:R2} should be obvious. This also explains why $\kappa_N^{n,m}(u,v)$ is called the prekernel. The main point is that if we know the weight function, $w_n^m(z)$, and the prekernel, $\kappa_N^{n,m}(u,v)$, then we know all the correlation functions.

It only remains to determine the skew-orthogonal polynomials and the constants $h_k^{n,m}$. For a general weight function the skew-orthogonal polynomials may be written as
\begin{subequations}
\begin{align}
p_{2k}^{n,m}(z)&=\average[\Big]{\prod_{i=1}^k (z-z_i)(z-z_i^\ast)}_{Z_k^{n,m}},\\
p_{2k+1}^{n,m}(z)&=\average[\Big]{\Big(z-\sum_{j=1}^k(z-z_j)\Big)\prod_{i=1}^k (z-z_i)(z-z_i^\ast)}_{Z_k^{n,m}},
\end{align}
\end{subequations}
which holds in complete analogue to~\cite{Kanzieper:2002}. It follows that the skew-orthogonal polynomials have real coefficients, $p_k(z)^\ast=p_k(z^\ast)$. In order to determine an explicit expression for the skew-orthogonal polynomials, we first notice that the weight function~\eqref{jpdf:weight2} is invariant under rotation in the complex plane, hence $w_n^m(z)=w_n^m(\abs z)$. It is well-known that monomials are orthogonal with respect to such weights, i.e. we have the relation
\begin{equation}
\int_\C d^2z\, w_n^m(z)\, z^k z^{\ast\ell}=s_k^{n,m} \delta_{k\ell},\quad\text{with}\quad 
s_k^{n,m}\equiv 2\pi \int_0^\infty dr\, w_n^m(r)\, r^{2k+1}.
\label{sop:orel}
\end{equation}
Using this relation it is straightforward to check that condition~\eqref{sop:rel1} is satisfied if
\begin{equation}
p_{2k+1}^{n,m}(z)=z^{2k+1}
\quad\text{and}\quad
p_{2k}^{n,m}(z)=\sum_{j=1}^k c_{2j}^{n,m} z^{2j}.
\end{equation}
Furthermore, condition~\eqref{sop:rel2} tells us that for any rotational invariant weight function, the constants $c_{2j}^{n,m}$ are given by a simple recursion relation,
\begin{equation}
c_{2j}^{n,m}=c_{2j-2}^{n,m}\frac{\int_0^\infty dr\, w_n^m(r)\, r^{4j-1}}{\int_0^\infty dr\, w_n^m(r)\, r^{4j+1}}.
\end{equation}
In the specific case where the weight function is given by equation~\eqref{jpdf:weight2}, the integrals can be performed analytically (see~\cite{GR:2000}) and the skew-orthogonal polynomials are in monic normalization given by
\begin{equation}
p_{2k+1}^{n,m}(z)=z^{2k+1}
\quad\text{and}\quad
p_{2k}^{n,m}(z)=\sum_{\ell=0}^k \bigg[\prod_{j=\ell+1}^{k} (m+2j)^n\bigg] z^{2\ell}.
\label{sop:poly}
\end{equation}
With this choice of polynomials we have
\begin{equation}
\inner{p_{2k+1}^{n,m}\vert p_{2\ell}^{n,m}}_S = h_k^{n,m}\delta_{k\ell}
\quad\text{with}\quad
h_k^{n,m}=\tfrac12(\pi\Gamma[m+2k+2])^n.
\label{sop:h}
\end{equation}
Inserting the skew-orthogonal polynomials~\eqref{sop:poly} and the constants $h_k^{n,m}$ into equation~\eqref{sop:ker1} we may write an explicit expression for the prekernel,
\begin{equation}
\kappa_N^{n,m}(u,v)=\frac{2}{(\pi\Gamma[m+1])^n}\sum_{k=0}^{N-1}\sum_{\ell=0}^k
\bigg[\frac{u^{2k+1}}{\prod_{j=0}^k(m+2j+1)^n}\frac{v^{2\ell}}{\prod_{j=1}^\ell(m+2j)^n}-(u\leftrightarrow v)\bigg].
\label{sop:ker2}
\end{equation}
Note that the prekernel agrees with known results for $n=1$~\cite{Kanzieper:2002} and $n=2$~\cite{Akemann:2005}. 

\begin{figure}[htp]
\centering
\includegraphics{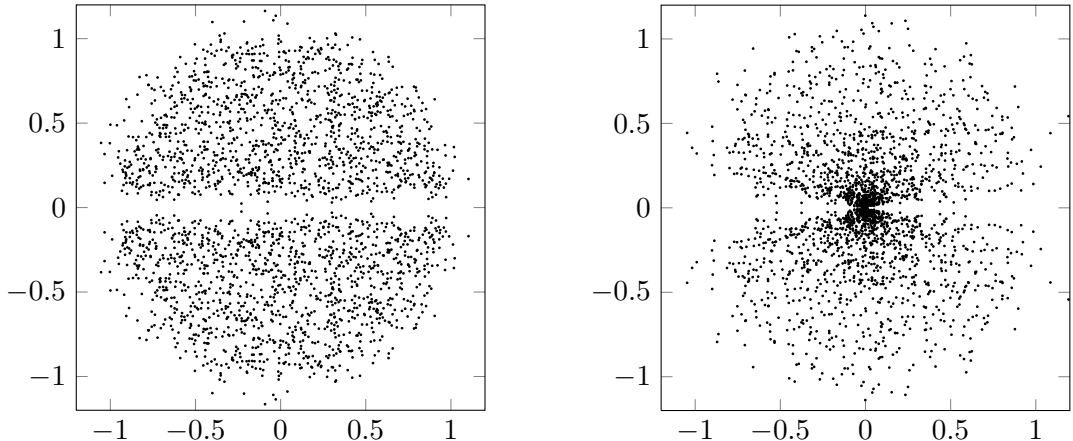}
\caption{Scatter plots of $50$ matrices defined as the product matrix, $P_n^m$, with $N=25$ and $m=0$, for $n=1$ (left) and $n=3$ (right). The eigenvalues have been rescaled by a factor $(2N)^{n/2}$ such that they approximately lie within the unit circle. Note that, due to the fact that eigenvalues come in complex conjugate pairs, the repulsion between individual eigenvalues results in a repulsion from the real axis.}
\label{fig:scat}
\end{figure}

Due to formula~\eqref{sop:R1} we know all eigenvalue correlations, since we have explicit expressions for both the weight function~\eqref{jpdf:weight2} and the prekernel~\eqref{sop:ker2}. As an example, the density of states, or one-point correlation function, is given in terms of~\eqref{jpdf:weight2} and~\eqref{sop:ker2} as
\begin{equation}
R_1^{n,m}(z)=(z^\ast-z)w_n^m(z)\kappa_N^{n,m}(z,z^\ast).
\label{sop:dos}
\end{equation}
Note that the first term on the right hand side of~\eqref{sop:dos} is responsible for a repulsion from the real axis. This repulsion can also be seen on figure~\ref{fig:scat}, which shows scatter plots of the product matrix~\eqref{intro:product}.

\section{Radial Behavior for the Density of States}
\label{sec:radial}

The radial behavior for the density of states for the product matrix, $P_n^m$, is an important quantity for several reasons. This is true even though the density of states is not rotational invariant (see figure~\ref{fig:scat}). The radial behavior is important for quantities such as hole probability and overcrowding at the origin, since these quantities are related to the radial ordering of the eigenvalues. Moreover, the radial behavior is often useful in comparison with numerical data. Therefore, in this section and section~\ref{sec:micro} we will mainly focus on the large-$N$ behavior of the radial density of states.

Note that studying the microscopic limits of the eigenvalue correlation functions, $R_k^{n,m}(z)$, in the complex plane for quaternion ($\beta=4$) Ginibre matrices is a challenging task; discussions of the microscopic origin limit have been given for $n=1$ and $n=2$ in~\cite{Kanzieper:2002} and~\cite{Akemann:2005}, respectively, but $n\geq 3$ remains an open problem. Averaging over the complex phase simplifies the formulae from the previous section considerably, which allows us to compare the results obtained in this paper directly to the results obtained for the product of complex ($\beta=2$) Ginibre matrices in~\cite{AB:2012}. 

We define the radial density of states, $\rho_N^{n,m}(r)$, as the average over the complex phase, i.e. 
\begin{equation}
\rho_N^{n,m}(r)=\frac{1}{2\pi}\int_{-\pi}^\pi d\theta\, R_1^{n,m}(re^{i\theta}),
\label{radial:rho1}
\end{equation}
where density of states, $R_1^{n,m}(z)$, is given by equation~\eqref{sop:dos} with weight function~\eqref{jpdf:weight2} and the prekernel~\eqref{sop:ker2}. We can use the orthogonality relation~\eqref{sop:orel} to perform the integration over $\theta$ in~\eqref{radial:rho1}, which yields
\begin{equation}
\rho_N^{n,m}(r)=\frac{2}{\pi}\,\MeijerG{n}{0}{0}{n}{-}{m,\ldots,m}{r^2} \sum_{k=0}^{N-1} \frac{r^{4k+2}}{\Gamma[m+2k+2]^n}.
\label{radial:rho2}
\end{equation}
Note that the density of states is normalized to the number of eigenvalues rather than to unity,
\begin{equation}
\int_0^\infty dr\, 2\pi r\,\rho_N^{n,m}(r)=\int_\C d^2z\, R_1^{n,m}(z)=2N.
\end{equation}
In the following we will discuss the large-$N$ behavior of radial density of states. In the limit of large matrices, the eigenvalues are distributed within a circle with radius $r\approx (m+2N)^{n/2}$. We will first discuss the macroscopic density of states, then we will turn to the discussion of the microscopic behavior at the origin and at the edges as well as in the bulk of the spectrum. 

The explicit expression for the radial density of states~\eqref{radial:rho2} for large $N$ and $r$ can be evaluated using the saddle point approach. A discussion of this is given in appendix~\ref{sec:asymp}. A similar discussion for the product of complex ($\beta=2$) Ginibre matrices was given in~\cite{AB:2012}. In the following we will also take $m$ to be proportional to the size of the product matrix, i.e. we take $m=2N\hat m$. For $N\gg 1$ and $m^{n/2}\lessapprox r \lessapprox (m+2N)^{n/2}$ we have (see appendix~\ref{sec:asymp})
\begin{equation}
\rho_{N}^{n,m}(r)\approx \frac{r^{\frac2n-2}}{n\pi}\,\frac{1}{2}
\left(\erfc\bigg[\sqrt{\frac{n}{2}}\frac{r^{2/n}-(m+2N)}{r^{1/n}}\bigg]
-\erfc\bigg[\sqrt{\frac{n}{2}}\frac{r^{2/n}-m}{r^{1/n}}\bigg]\right).
\label{radial:rho3}
\end{equation}
This structure leads us to introduce a scaled density of states defined as
\begin{equation}
\hat\rho_{N}^{n,m}(\hat r)\equiv (2N)^{n-1}\rho_N^{n,m}(\hat r(2N)^{n/2}).
\label{radial:scale}
\end{equation}
We rescale of the radial variable, $r$, in order to get compact support, while the prefactor ensures that the density is finite in the limit of large matrices. Using the asymptotic behavior~\eqref{radial:rho3}, the scaled density of states~\eqref{radial:scale} can be written as
\begin{equation}
\hat\rho_{N}^{n,m}(\hat r)\approx \frac{\hat r^{\frac2n-2}}{n\pi}\,\frac{1}{2}
\left(\erfc\bigg[\sqrt{nN}\frac{\hat r^{2/n}-(\hat m+1)}{\hat r^{1/n}}\bigg]
-\erfc\bigg[\sqrt{nN}\frac{\hat r^{2/n}-\hat m}{\hat r^{1/n}}\bigg]\right).
\end{equation}
The complementary error functions change only in the neighborhood of $\hat r=\hat m^{n/2}$ and $\hat r=(\hat m+1)^{n/2}$, respectively. Taylor expanding the arguments of the error functions around these values we obtain
\begin{equation}
\hat\rho_{N}^{n,m}(\hat r)\approx \frac{\hat r^{\frac2n-2}}{n\pi}\,\frac{1}{2}
\left(\erfc\bigg[\sqrt{\frac{4N}{n}}\frac{\hat r-(\hat m+1)^{n/2}}{(\hat m+1)^{(n-1)/2}}\bigg]
-\erfc\bigg[\sqrt{\frac{4N}{n}}\frac{\hat r-\hat m^{n/2}}{\hat m^{(n-1)/2}}\bigg]\right).
\label{radial:rho4}
\end{equation}
As a final remark in this section, we will determine the macroscopic (or mean) density of states. The macroscopic radial density of states is given by
\begin{equation}
\hat \rho_\macro^{n,m}(\hat r)\equiv \lim_{N\to\infty}\hat\rho_{N}^{n,m}(\hat r)=
\frac{\hat r^{\frac2n-2}}{n\pi}\big(\theta[(\hat m+1)^{n/2}-\hat r]-\theta[m^{n/2}-\hat r]\big),
\label{radial:macro}
\end{equation}
where $\theta$ denotes the Heaviside step function. Note that that macroscopic radial density of states for the product of $n$ quaternion ($\beta=4$) Ginibre matrices~\eqref{radial:macro} is similar to that complex ($\beta=2$) Ginibre matrices, see~\cite{BJW:2010,GT:2010,RS:2011,AB:2012}. The induced Ginibre ensemble has also been discussed for a single real ($\beta=1$) matrix and a single complex ($\beta=2$) matrix~\cite{FBKSZ:2012}, which also gives a structure similar to~\eqref{radial:macro} with $n=1$. The macroscopic radial density of states~\eqref{radial:macro} tells us that the eigenvalues of the product of induced Ginibre matrices lie within an annulus centered at origin in the complex plane. That this is an important generalization is illustrated by the single ring theorem~\cite{FZ:1997,GKZ:2011}.
\begin{figure}[htp]
\centering
\includegraphics{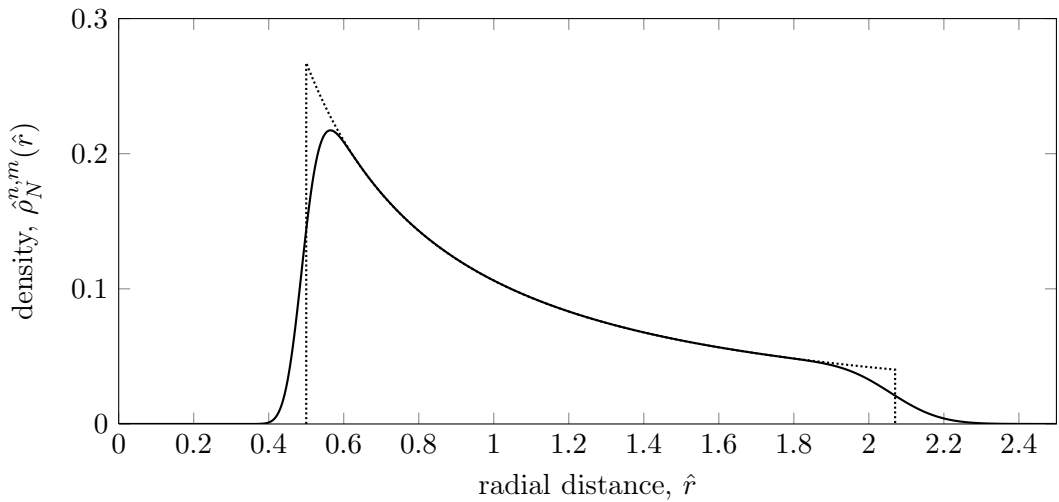}
\caption{Plot for the radial density of states, $\hat\rho_N^{n,m}(\hat r)$, with $N=100$ and $m=125$ for $n=3$. The dotted curve indicates the macroscopic limit.}
\label{fig:macro}
\end{figure}

\section{Microscopic Limits and Universality}
\label{sec:micro}

In this section we will discuss microscopic limits of the radial density of states. The asymptotic formulae of previous section enable us to identify the different regions where we can take the microscopic limit. We will discuss the universal bulk limit as well as the universality at the inner and outer edge, and finally we discuss the microscopic origin limit.

\paragraph{Universal Bulk Limit} The microscopic bulk limit can be taken, when the radial variable is far away from the edges, $m^{n/2}\ll r\ll (m+2N)^{n/2}$. In order to study the microscopic behavior we must rescale the variable $r\to r(2N)^{-n/2}$ to get compact support, and then make another rescaling $r\to r(2N)^{n/2}$ to get to the level of the eigenvalue fluctuations. Clearly, the two scalings cancel each other. The final step to obtain the microscopic bulk limit is to unfold the variable, such that the density of states becomes locally flat. From previous section we know that the unfolding can be made with a change of variable given by $\tilde r=r^{2/n}$. In the region of interest, $m^{n/2}\ll r\ll (m+2N)^{n/2}$, this means that 
\begin{equation}
d\tilde r=2\pi r\Big[\lim_{N\to\infty} \rho_{N}^{n,m}(r)\Big]dr,
\end{equation}
which indeed gives a flat density. We get a microscopic bulk radial density given by
\begin{equation}
\rho_\bulk^{n,m}(\tilde r)=\frac{1}{2\pi\tilde r}.
\end{equation}
Note that the microscopic bulk limit of the radial density of states is independent of $n$ and $m$. Moreover, the limit is also equivalent to the microscopic bulk limit of the radial density of states for the product of complex ($\beta=2$) Ginibre matrices~\cite{AB:2012}. Of course the main interest in the bulk is the higher order correlation functions in the complex plane, $R_k^{n,m}(z)$.

\paragraph{Universal Edge Limit} We will now investigate the microscopic behavior at the inner and outer edge, i.e. for $r\approx m^{n/2}$ and $r\approx (m+2N)^{n/2}$, respectively. Looking at equation~\eqref{radial:rho4} from previous section, we introduce the edge variables, $\hat\epsilon_\text{in}$ and $\hat\epsilon_\text{out}$, given by
\begin{equation}
\hat\epsilon_\text{in} \equiv -\sqrt{\frac{2N}{n}} \, \frac{\hat r-\hat m^{n/2}}{\hat m^{(n-1)/2}} \quad\text{and}\quad
\hat\epsilon_\text{out}\equiv  \sqrt{\frac{2N}{n}} \, \frac{\hat r-(\hat m+1)^{n/2}}{(\hat m+1)^{(n-1)/2}},
\label{micro:epsilon}
\end{equation}
with the scaled quantities $\hat r$ and $\hat m$ defined in previous section. The error functions changes only in a narrow region, and the width of this region is proportional to $1/\sqrt{2N}$ when we use the scaled variable $\hat r$. With this in mind, it is clear that the edge variables~\eqref{micro:epsilon} zoom in on the region of change (for the inner or outer edge, respectively) and then rescale such that the edge is located at unity.

With the above definition of the edge variables~\eqref{micro:epsilon} we get the universal edge behavior
\begin{equation}
\hat\rho_\edge^{n,m}(\hat\epsilon)=\lim_{N\to\infty}\hat\rho_N^{n,m}(\hat\epsilon)
=\frac{1}{2\pi}\erfc[\sqrt{2}\,\hat\epsilon],
\end{equation}
where $\hat\epsilon$ is defined as $\hat\epsilon_\text{in}$ and $\hat\epsilon_\text{out}$ for the inner and outer edge, respectively. The result holds for any $n$ and $\hat m$ and is thus universal; see also the discussions given in~\cite{Mehta:2004,KS:2010,FBKSZ:2012,AB:2012}. It should be mentioned that if $m$ is small, i.e. $m$ is not proportional to the size of the matrix, then there is no inner edge. Instead there exists a microscopic origin limit which does depend on both $n$ and $m$; we will discuss this limit below.

\paragraph{Microscopic Origin Limit} We have already seen that when $m\gg1$, then the eigenvalues are repulsed from the origin. Here we will look at the situation where $m$ is of order unity, and discuss the microscopic behavior at the origin. The microscopic limit can be obtained by taking $N\to\infty$ while keeping the radial variable far away from the edge, $r\ll (2N)^{n/2}$. If we do so directly in equation~\eqref{radial:rho2} we get
\begin{equation}
\rho_\origin^{n,m}(r)\equiv\lim_{N\to\infty}\rho_N^{n,m}(r)
=\frac{2}{\pi}\,\MeijerG{n}{0}{0}{n}{-}{m,\ldots,m}{r^2} \sum_{k=0}^\infty \frac{r^{4k+2}}{\Gamma[m+2k+2]^n}.
\end{equation}
The infinite series can be recognized as a generalized hypergeometric function, and we can write the microscopic origin limit as
\begin{multline}
\quad\rho_\origin^{n,m}(r)=\frac{2r^2}{\pi\Gamma[m+2]}\,\MeijerG{n}{0}{0}{n}{-}{m,\ldots,m}{r^2} \\
\times\hypergeometric{1}{2n}{1}{\frac{m+2}{2},\ldots,\frac{m+2}{2},\frac{m+3}{2},\ldots,\frac{m+3}{2}}{\frac{r^4}{2^{2n}}},\quad
\end{multline}
where both $(m+2)/2$ and $(m+3)/2$ appear $n$ times in the hypergeometric function. Note that the microscopic origin limit depends on both $n$ and $m$. This is to be compared with the microscopic limit for the product of complex ($\beta=4$) Ginibre matrices given in~\cite{AB:2012}.

\section{Conclusions and Outlook}

In this paper we have investigated the spectral properties for the product of $n$ independent induced quaternion ($\beta=4$) Ginibre matrices. We have shown that the joint probability density function for the product matrix is identical to that for a single Ginibre matrix, except that the weight function is more complicated. Furthermore, we have seen that the weight function for the product quaternion ($\beta=4$) Ginibre matrices is identical to that for the product of complex ($\beta=2$) Ginibre matrices, i.e. is given in terms of a Meijer $G$-function.

For matrices of finite size, we have used the method of skew-orthogonal polynomials to obtain an explicit expression for the prekernel. This implies that we can write down any correlation function of a product matrix with finite size, since all eigenvalue correlations can be expressed in terms of weight function and the prekernel. 

Finally, we looked at the radial density of states, defined from the density of states in the complex plane by averaging over the complex phase. In this case we investigated the limit of large matrices. We found that the macroscopic radial density was similar to the radial behavior for the product of complex ($\beta=2$) Ginibre matrices. Furthermore, we discussed the microscopic limits. For the weakly induced ensemble, $m=\cO(N^0)$, we studied the microscopic origin limit, while for the strongly induced ensemble, $m\propto N$, we discussed the universal behavior at the inner and outer edge. We also looked at the behavior in the bulk. We saw that the edge behavior, given in term of a complementary error function, was universal and identical to the case of complex Ginibre matrices. The microscopic origin limit was written using a generalized hypergeometric function.

It would be a natural extension of this paper to study the large matrix limit of the correlation functions in the complex plane, i.e. without taking the average over the complex phase. 
Furthermore, it is an appealing challenge to extend the discussion of products of Ginibre matrices to include the case of real ($\beta=1$) Ginibre matrices. We expect that the results for the edge and the bulk to hold for the complex eigenvalues, while the origin will be generalized as already known for the product of two matrices. It would also be interesting to discuss the hole probability and overcrowding at the origin for quaternion matrices, this was done for the product of complex matrices in~\cite{AS:2012}. Diagrammatic methods have previously been used to discuss the macroscopic limit for the product of complex rectangular matrices~\cite{BJLNS:2010} and it would be desirable to understand this on the level of orthogonal polynomials. Several of these projects are currently on the way.

\paragraph{Acknowledgment} G. Akemann, M. Kieburg, E. Strahov as well as the participants at the `VIII Brunel--Bielefeld Workshop on Random Matrix Theory' are thanked for useful discussions. The author is supported by the German Science Foundation (DFG) through the International Graduate College `Stochastics and Real World Models' (IRTG 1132) at Bielefeld University.

\appendix

\section{Derivation of the Jacobian}
\label{sec:jacob}

In this appendix we calculate the Jacobian for the change of variables corresponding to the parameterization given in section~\eqref{sec:jpdf}, i.e. the the change of variables from $X_i$ to $\Lambda_i$, $T_i$ and $U_i$. Recall that $X_i$ are quaternion ($\beta=4$) Ginibre matrices, $\Lambda_i$ are diagonal  (block) matrices, $T_i$ are strictly upper-triangular (block) matrices and $U_i\in Sp(N)/U(1)^N$ are symplectic matrices.

In the following $dX=[dX_{ab}]$ denotes one-forms ordered as a matrix, while $DX=\bigwedge dX_{ab}$ denotes the corresponding exterior product of independent one-forms, and $\abs{DX}$ is the unoriented volume element. The one-forms of $dX_i$ are related to those of $d\Lambda_i$, $dT_i$ and $dU_i$ as
\begin{equation}
dX_i=U_idY_iU_{i+1}^{-1},
\end{equation}
where
\begin{equation}
dY_i=d\Lambda_i+dT_i+dM_i,
\label{jacob:dY}
\end{equation}
with $dM_i$ defined as
\begin{equation}
(dM_i)_{ab}=(dA_i)_{ab}(\Lambda_i)_{bb}-(\Lambda_i)_{aa}(dA_{i+1})_{ab}+\sum_{c=1}^N(dA_i)_{ac}(T_i)_{cb}-\sum_{c=1}^N(T_i)_{ac}(dA_{i+1})_{cb}.
\label{jacob:dM}
\end{equation}
Here $dA_{n+1}=dA_1$ and $dA_{i}=U_i^{-1}dU_i$ are anti-self-dual matrices~\cite{Mehta:2004}. Due to the anti-self-duality the one-forms in the upper and lower triangular entries are related by $(dA_i)_{ab}=-(d\overline{A_i})_{ba}$, where the bar denotes the quaternion conjugate~\cite{Mehta:2004}. For this reason we will only keep the lower triangle as independent one-forms. Furthermore, it should be noted that since $U_i\in Sp(N)/U(1)^N$ the diagonal entries $(dA_i)_{aa}$ has two independent one-forms (each entry in the lower triangle has four independent one-forms),
\begin{equation}
(dA_i)_{aa}=\qj (dA_i^{(0)})_{aa}+\qk (dA_i^{(1)})_{aa}
=\frac{1}{2} \begin{pmatrix} 0 & -d\alpha_{ia}^\ast \\ d\alpha_{ia} & 0 \end{pmatrix},
\end{equation}
where $\alpha_{ia}=-(A_i^{(2)})_{aa}+i(A_i^{(3)})_{aa}\in\C$.

We can now calculate the Jacobian. It is trivially seen that the Jacobian for the change of variables from $X$ to $Y$ equals one, since $U_i$ are symplectic matrices. In order to find the Jacobian for the change variables from $Y$ to $\Lambda$, $T$ and $M$ we write~\eqref{jacob:dY} as
\begin{alignat}{6}
(dY_i^{(q)})_{ab}&=(d\Lambda_i^{(q)})_{ab}&		&			&+\ph{(dM_i^{(q)})_{ab}}&(dM_i^{(q)})_{ab}	&\quad\text{for }a=b\text{ and }q=0,1, \nn\\
(dY_i^{(q)})_{ab}&=			&(dT_i^{(q)})_{ab}&			&+(dM_i^{(q)})_{ab}	&			&\quad\text{for }a<b\text{ and }q=0,1, \nn\\
(dY_i^{(q)})_{ab}&=			&		&(dT_i^{(q)})_{ab}	&+\ph{(dM_i^{(q)})_{ab}}&(dM_i^{(q)})_{ab}	&\quad\text{for }a<b\text{ and }q=2,3, \nn\\
(dY_i^{(q)})_{ab}&=			&		&			&+(dM_i^{(q)})_{ab}	&			&\quad\text{for }a>b\text{ and }q=0,1, \nn\\
(dY_i^{(q)})_{ab}&=			&		&			&			&(dM_i^{(q)})_{ab}	&\quad\text{for }a\geq b\text{ and }q=2,3.
\end{alignat}
With this ordering it is immediately seen that the Jacobian matrix for the change of variables $Y$ to $\Lambda$, $T$ and $M$ is upper triangular and that the Jacobian is equal to unity, such that
\begin{equation}
\abs{DX}=\abs{DY}=\abs{D\Lambda}\abs{DT}\abs{DM}.
\end{equation}
Here $D\Lambda$ and $DT$ are the Euclidean volume elements for $\Lambda$ and $T$, respectively. The measure $DM$ is given by
\begin{equation}
DM=\Big(\bigwedge_{a>b}^N\bigwedge_{q=0}^3\bigwedge_{i=1}^n (dM_i^{(q)})_{ab}\Big)
\wedge\Big(\bigwedge_{b=1}^N\bigwedge_{q=0}^1\bigwedge_{i=1}^n (dM_i^{(q)})_{bb}\Big).
\label{jacob:DM}
\end{equation}
Note that the volume elements $DM$ and $DA$ contain the same number of independent one-forms, which was to be expected from~\eqref{jacob:dM}. As the final step we want change variables from $M$ to $A$. First we will show that the Jacobian matrix $J=\p M/\p A$ can be chosen to be triangular and that the Jacobian $\abs J$ is independent of $T$. To do so, we order the matrix index $\alpha=ab$ such that $ab<a'b'$ if $a>a'$ or if $a=a'$ and $b<b'$. This is identical to the ordering used in~\cite{AB:2012}. The one-forms $dM$ and $dA$ are related by
\begin{equation}
(dM_i)_\alpha=\sum_{j,\alpha'} (J_{ij})_{\alpha\alpha'}(dA_j)_{\alpha'},
\end{equation}
where $J$ is the Jacobian matrix. Using that the matrices $T_i$ are upper triangular, we see from~\eqref{jacob:dM} that the entries with $\alpha>\alpha'$ in the Jacobian matrix are zero, i.e. the Jacobian matrix is triangular such that only the diagonal entries ($\alpha=\alpha'$) contribute to the Jacobian, $\abs J$. It is clear from~\eqref{jacob:dM} that the diagonal entries do not depend on $T$, and we can therefore replace the $DM$ with
\begin{equation}
(d\widehat M_i)_{ab}=(dA_i)_{ab}(\Lambda_i)_{bb}-(\Lambda_i)_{aa}(dA_{i+1})_{ab}.
\label{jacob:dMhat}
\end{equation}
We use~\eqref{jacob:dMhat} in~\eqref{jacob:DM} in order to obtain the final Jacobian. Recall that the entries $(d\widehat M_i)_{ab}$ and $(dA_i)_{ab}$ are quaternions, or equivalently $2\times 2$ matrices. For $a>b$ we have that
\begin{align}
(d\widehat M_i)_{ab}&=
\begin{pmatrix}
du_i & -dv_i^\ast \\
dv_i & du_i^\ast
\end{pmatrix}_{ab}
\begin{pmatrix}
x_i & 0 \\
0 & x_i^\ast
\end{pmatrix}_{bb}
-
\begin{pmatrix}
x_i & 0 \\
0 & x_i^\ast
\end{pmatrix}_{aa}
\begin{pmatrix}
du_{i+1} & -dv_{i+1}^\ast \\
dv_{i+1} & du_{i+1}^\ast
\end{pmatrix}_{ab} 		\nn\\
&=
\begin{pmatrix}
x_{ib}du_{iab}-x_{ia}du_{(i+1)ab} & -x_{ib}^\ast dv_{iab}^\ast+x_{ia}dv_{(i+1)ab}^\ast \\
x_{ib}dv_{iab}-x_{ia}^\ast dv_{(i+1)ab} & x_{ib}^\ast du_{iab}^\ast -x_{ia}^\ast du_{(i+1)ab}^\ast 
\end{pmatrix}.
\end{align}
A similar equation holds for $a=b$ but in that case there are no diagonal terms. Instead of using the quaternion index $(q)$, we will use the matrix index $k\ell$, i.e. $(d\widehat M_i)_{ab}^{k\ell}$ with $k,\ell=1,2$ and similarly for $dA$. We see that
\begin{align}
(a>b):\quad \bigwedge_{i=1}^n(d\widehat M_i)_{ab}^{11} &= (x_{1b}x_{2b}\cdots x_{nb}-x_{1a}x_{2a}\cdots x_{na})\bigwedge_{i=1}^n du_{iab}, \nn\\
(a>b):\quad \bigwedge_{i=1}^n(d\widehat M_i)_{ab}^{21} &= (x_{1b}x_{2b}\cdots x_{nb}-x_{1a}^\ast x_{2a}^\ast \cdots x_{na}^\ast )\bigwedge_{i=1}^n dv_{iab},\quad\text{etc.}
\end{align}
Combining these results we find the Jacobian
\begin{equation}
\abs{J(z)}=\abs[\bigg]{\frac{\p\widehat M(z)}{\p A}}
=\prod_{a>b}^N\, \abs{z_{b}-z_{a}}^2 \abs{z_{b}-z_{a}^\ast}^2 \prod_{c=1}^N\, \abs{z_{c}-z_{c}^\ast}^2,
\end{equation}
where $z$ is a short hand notation for the product $z_a=\prod_{i=1}^n x_{ia}$. This is the Jacobian given in section~\ref{sec:jpdf}.

\section{Asymptotic of Radial Density of States}
\label{sec:asymp}

In this appendix we discuss the asymptotic behavior of the radial density of states. We study the asymptotics for large matrices ($N\gg1$) and large $r$ (this means $1\ll r\lessapprox (2N)^{n/2}$ if $m$ is of order unity and $m^{n/2}\lessapprox r\lessapprox (m+2N)^{n/2}$ if $m\gg1$). Recall that the radial density of states is given by
\begin{equation}
\rho_N^{n,m}(r)=\frac{2}{\pi}\,\MeijerG{n}{0}{0}{n}{-}{m,\ldots,m}{r^2} \sum_{k=0}^{N-1} \frac{r^{4k+2}}{\Gamma[m+2k+2]^n}.
\label{asymp:rho}
\end{equation}
The asymptotics for the Meijer $G$-function at large arguments is known, see~\cite{Fields:1972}. Furthermore, the particular Meijer $G$-function given in~\eqref{asymp:rho} has been discussed explicitly in~\cite{AB:2012}. For this reason we refer the reader to the above mentioned references and state the result without derivation,
\begin{equation}
\MeijerG{n}{0}{0}{n}{-}{m,\ldots,m}{r^2} \approx \frac{(2\pi)^{(n-1)/2}}{\sqrt n}\, r^{(1-n)/n}\, r^{2m}\, e^{-nr^{2/n}}
\quad\text{for}\quad r\gg 1.
\label{asymp:G}
\end{equation}
In order to evaluate the sum in~\eqref{asymp:rho} we will use a saddle point approximation. For large $N$ the sum can be replaced by an integral. If we also make a shift in the integration, $k\to (k-m-1)/2$, we see that
\begin{equation}
\sum_{k=0}^{N-1} \frac{r^{4k+2}}{\Gamma[m+2k+2]^n}\approx
\frac{1}{2}\int_{m}^{m+2N}dk\, \frac{r^{2k-2m}}{\Gamma[k+1]^n}
\end{equation}
The gamma function is well approximated by Stirling's formula, $\Gamma[k+1]\approx \sqrt{2\pi k}(k/e)^k$. After this replacement we have
\begin{equation}
\sum_{k=0}^{N-1} \frac{r^{4k+2}}{\Gamma[m+2k+2]^n}\approx
\frac{(2\pi)^{-n/2}}{2}r^{-2m}\int_{m}^{m+2N}dk\,k^{-n/2}\,e^{S(k)},
\label{asymp:sum1}
\end{equation}
where
\begin{equation}
S(k)=k\log r^2-nk\log k+nk.
\end{equation}
Note that a similar integral also appeared in the evaluation of the kernel for the product of complex ($\beta=2$) Ginibre matrices~\cite{AB:2012}. In order to evaluate the integral we approximate the exponential in~\eqref{asymp:sum1} by a Gaussian function. The solution of the saddle point equation,
\begin{equation}
S\,'(k_\ast)=\log r^2-n\log k_\ast=0,
\end{equation}
gives the location of the maximum of the Gaussian function, $k_\ast=r^{2/n}$. At the maximum we have $S(k_\ast)=nk_\ast$ and $S\,''(k_\ast)=-n/k_\ast$, and therefore we have that
\begin{equation}
\sum_{k=0}^{N-1} \frac{r^{4k+2}}{\Gamma[m+2k+2]^n}\approx
\frac{(2\pi)^{-n/2}}{2}r^{-2m}e^{nk_\ast}\int_{m}^{m+2N}dk\,k_\ast^{-n/2}\,e^{-n(k-k_\ast)^2/2k_\ast}.
\label{asymp:approx}
\end{equation}
We have $r\gg 1$ such that the integrand is a narrow peak centered at $k_\ast$. For this reason we can replace the integration limits with $\pm\infty$ if $m\ll k_\ast\ll m+2N$. In this region it straight forward to perform the integral,
\begin{equation}
\sum_{k=0}^{N-1} \frac{r^{4k+2}}{\Gamma[m+2k+2]^n}\approx
\frac{(2\pi)^{(1-n)/2}}{2}r^{-2m}r^{(1-n)/n}e^{nr^{2/n}}.
\end{equation}
Inserting this result and the asymptotic behavior of the Meijer $G$-function~\eqref{asymp:G} into~\eqref{asymp:rho}, we obtain
\begin{equation}
\rho_{N}^{n,m}(r)\approx \frac{r^{\frac2n-2}}{n\pi}
\quad\text{for}\quad
m^{n/2}\ll r\ll (m+2N)^{n/2}.
\end{equation}
This gives the large-$N$ asymptotic of the radial density of states in the bulk, but if we want to study the behavior at the edges we need to be more careful. If the maximum of the Gaussian peak is in the neighborhood of the edges of the integration interval, $k_\ast\approx m$ and $k_\ast\approx m+2N$, then we cannot replace the integration limits with $\pm\infty$. Instead we must consider situation where the Gaussian peak lies partly outside the integration interval. If we write~\eqref{asymp:approx} as
\begin{equation}
\sum_{k=0}^{N-1} \frac{r^{4k+2}}{\Gamma[m+2k+2]^n}\approx
\frac{(2\pi)^{-n/2}}{2}r^{-2m}e^{nk_\ast}\bigg(\int_{-\infty}^{m+2N}dk-\int_{-\infty}^{m}dk\Bigg)\,k_\ast^{-n/2}\,e^{-n(k-k_\ast)^2/2k_\ast},
\end{equation}
it is clear that the evaluation of the integral at the inner and outer edge is identical, since the Gaussian peak is symmetric around its maximum. Partly integrating over the Gaussian peak results in a complementary error function. Combining this with the discussion above we obtain the following asymptotic behavior of the radial density of states,
\begin{equation}
\rho_{N}^{n,m}(r)\approx \frac{r^{\frac2n-2}}{n\pi}\,\frac{1}{2}
\left(\erfc\bigg[\sqrt{\frac{n}{2}}\frac{r^{2/n}-(m+2N)}{r^{1/n}}\bigg]
-\erfc\bigg[\sqrt{\frac{n}{2}}\frac{r^{2/n}-m}{r^{1/n}}\bigg]\right).
\end{equation}
If $m$ is small ($m\sim1\ll N$), then we only have an outer edge and the density of states becomes
\begin{equation}
\rho_{N}^{n,m}(r)\approx \frac{r^{\frac2n-2}}{n\pi}\,\frac{1}{2}
\erfc\bigg[\sqrt{\frac{n}{2}}\frac{r^{2/n}-2N}{r^{1/n}}\bigg],
\end{equation}
which holds for $1\ll r\lessapprox 2N$.


\bibliographystyle{unsrt}
\bibliography{ref}

\end{document}